# On Cosmic-Ray-Driven Electron Reaction Mechanism for Ozone Hole and Chlorofluorocarbon Mechanism for Global Climate Change

Qing-Bin Lu

Department of Physics and Astronomy, University of Waterloo, Waterloo, ON, N2L 3G1, Canada.

E-mail: qblu@uwaterloo.ca

**Abstract**

Numerous laboratory measurements have provided a sound physical basis for the cosmic-ray driven electron-induced reaction (CRE) mechanism of halogen-containing molecules for the ozone hole. And observed spatial and time correlations between polar ozone loss or stratospheric cooling and cosmic rays have shown strong evidence of the CRE mechanism [Q.-B. Lu, Phys. Rep. **487**, 141-167(2010)]. Chlorofluorocarbons (CFCs) were also long-known greenhouse gases but were thought to play only a minor role in climate change. However, recent observations have shown evidence of the saturation in greenhouse effect of non-CFC gases. A new evaluation has shown that halocarbons alone (mainly CFCs) could account for the rise of 0.5~0.6 °C in global surface temperature since 1950, leading to the striking conclusion that not $CO_2$ but CFCs were the major culprit for global warming in the late half of the 20th century [Q.-B. Lu, J. Cosmology **8**, 1846-1862(2010)]. Surprizingly, a recent paper [J.-W. Grooß and R. Müller, Atmos. Environ. **45**, 3508-3514(2011)] has criticized these new findings by presenting "ACE-FTS satellite data". Here, I show that there exist serious problems with such "ACE-FTS satellite data" because the satellite has essentially not covered the Antarctic vortex in the presented months (especially winter months during which most effective CRE reactions are expected) and that the criticisms do not agree with the scientific facts in the literature. Instead, real data from multiple satellites provide strong evidence of the CRE mechanism. So far, the CRE mechanism is the only one that reproduces and predicts 11-year cyclic variations of ozone loss in the Antarctic $O_3$ hole and of resultant stratospheric cooling, and the CFC mechanism can well explain both recent global warming and cooling. These findings should improve our understandings of the ozone hole and global climate change.

**1. Introduction**
Both natural and human effects could alter the Earth's climate and environment. The ozone hole and global temperature change have been two major scientific problems of global concern. There is long interest in studying the effects of cosmic rays (CRs) on Earth's ozone layer [1-17]. In the 1970s, the odd nitrogen ($NO_x$) generated by solar particle events (SPEs) were proposed first by Crutzen et al. [1] for solar proton events and then by Thorne [3] for energetic electron precipitation events to cause transient $O_3$ destruction in the upper stratosphere at altitudes above 30 km. And Ruderman et al. [2] proposed that the 11-year solar cycle variation of the CR intensity may also result in a small modulation (2~3% above or below the mean value) of polar total $O_3$. However, the sink of $O_3$ by SPEs, often associated with very large solar flares, is expected to be most pronounced during solar maxima and opposite in phase to the $O_3$ loss caused by CRs [3]. If these natural effects were appreciable, they would lead to an 11-year cyclic variation in any season (e.g., summer). However, observed $O_3$ data show no considerable long-term correlation between total ozone in the *summer* polar stratosphere and solar activity / CRs [15]. These natural effects are very limited in the long-term total $O_3$ variation.

Direct measurements based on balloons and satellites have shown convincing evidence that the formation of the $O_3$ hole is related to human-made chlorofluorocarbons (CFCs) such as $CF_2Cl_2$ (CFC-12) and $CFCl_3$ (CFC-11). In 1974, Molina and Rowland [18] first proposed that CFCs are decomposed by photodissociation with UV sunlight (a process called photolysis). The liberated chlorine atoms contribute to the depletion of the $O_3$ layer. This photolysis was originally predicted to happen in the upper tropical stratosphere at high altitudes of ~40 km. Then it came with a surprising observation by Farman, Gardiner and Shanklin [19] in 1985 that the springtime $O_3$ hole appeared over Antarctica and at low altitudes of 15-20 km. It was subsequently found that the formation of the ozone hole is closely related to the existence of polar stratospheric clouds (PSCs) that form in the winter Antarctic stratosphere and consist mainly of condensed-phase water ice or/and nitric acid ice [20, 21]. The $O_3$ hole was then explained by mixed photochemical models [22-25]: (1) the photolysis of CFCs occurs in the upper tropical stratosphere; (2) air transportation to the lower polar stratosphere of inorganic halogen species (mainly HCl and $ClONO_2$) resulting from reactions of CFC dissociation products (Cl and ClO) with other atmospheric molecules ($CH_4$ and $NO_2$); (3) heterogeneous chemical reactions of inorganic halogen species on ice surfaces in PSCs to form photoactive $Cl_2$ and HOCl in the winter lower polar stratosphere. Finally, the sunlight-photolysis of photoactive halogens produces Cl atoms to destroy ozone in the spring polar stratosphere. These are the widely accepted explanation of the $O_3$ hole.

The Montreal Protocol has successfully phased out the production of CFCs in the world wide. Since the observed total halogen level in the troposphere peaked in ~1994, the original prediction was that "*Peak global ozone losses are expected to occur during the next several years*" [26]. The equivalent effective stratospheric chlorine levels at mid-latitudes and Antarctica were then re-calculated to peak in the years around 1997 and 2000, respectively with delays of ~3 and ~6 years from the tropospheric peak, and it was thus predicted that the total $O_3$ in mid-latitudes and the Antarctic $O_3$ hole would have recovered correspondingly [27]. So far, however, no statistically significant recovery of $O_3$ loss has been observed [28]. Even the largest Arctic ozone hole was observed in 2011 [29]. More remarkably, the largest (smallest) Antarctic $O_3$ holes were observed when solar activity was weakest (strongest), e.g., in 1987, 1998 and 2008 (1991, 2002 and 2013 (expected)). In fact, there has been no $O_3$ loss observed over the Equator in the past four decades. These observations are inconsistent with the above predictions from photochemical models and indicate that the current photochemical theory of ozone loss is incomplete or wrong. As noted recently by Manney et al. [29], the ability of current atmospheric/climate models to predict the future polar $O_3$ loss is very limited, and improving the predictive capabilities is one of the greatest challenges in polar $O_3$ research. To place the Protocol on a firmer scientific ground, it is still required to obtain a correct and complete ozone depletion theory.

The fact is also that parallel to the study of photolysis of CFCs, there is a long history of studying electron-induced reactions of halogenated molecules including CFCs [30, 31]. The dissociative attachment (DA) of gaseous CFCs to low-energy free electrons was once suggested as a potential sink of CFCs in the atmosphere by Peyerimhoff et al. [32, 33]. But the process was long thought to be insignificant due to the low free electron density detected in the stratosphere [34, 35]. Then, the large enhancements by *up to four orders of magnitude* in electron-stimulated desorption of $Cl^−$ ions from $CF_2Cl_2$ adsorbed on polar molecular ice surfaces were surprisingly observed by Lu and Madey [5, 36-39] and then confirmed by Solovev et al. [40]. In Lu and Madey experiments [5], electron-induced dissociation cross sections of CFCs adsorbed on polar ice surfaces were measured to be *$10^6$-$10^8$ times* the photodissociation cross sections ($10^{-20}$ cm$^2$) of gaseous CFCs [30], and a dissociative electron transfer (DET) mechanism was proposed to explain the results:

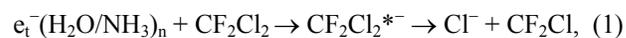
$e_t^−(H_2O/NH_3)_n + CF_2Cl_2 \rightarrow CF_2Cl_2^{*−} \rightarrow Cl^− + CF_2Cl$, (1)

where $e_t^−$ is a weakly-bound electron trapped in the polar ($H_2O/NH_3$) ice [5, 36]. This unexpected finding revived the studies of electron-induced reactions of halogenated molecules. The DET mechanism of halogen-containing molecules was also confirmed in surface electron trapping experiments by Lu and Sanche [6, 41-43] and in surface photochemistry experiments by others [44, 45]. More recently, femtosecond time-resolved laser spectroscopic measurements have obtained direct observations of DET reactions of halogenated molecules in liquid water by Lu and co-workers [46-49] or adsorbed on solid ice surfaces by Ryu et al. [50] and Wolf and co-workers [51, 52]. Remarkably, Stähler et al. [52] have recently measured a very large DET dissociation cross section up to $4\times10^{−12}$ cm$^2$ for $CFCl_3$ on $D_2O$ ice, which is comparable to those observed for $CF_2Cl_2$ adsorbed on $H_2O$ and $NH_3$ ice, being $\sim1\times10^{−14}$ and $\sim6\times10^{−12}$ cm$^2$, respectively by Lu and Madey [5]. The DET mechanism has also been confirmed by several theoretical simulations [53-57]. As reviewed recently by Lu [15], it has now been well-established that polar media in various (gas, liquid and solid) phases can largely enhance electron-induced dissociations of both organic and inorganic halogenated molecules such as CFCs and HCl (ClONO$_2$) to various degrees via the DET reaction mechanism.

It is also well-known that copious electrons are produced by atmospheric ionization of cosmic rays in the stratosphere, especially in the lower polar stratosphere with the presence of PSC ice particles in the winter and early spring polar stratosphere. This logically led to the search of the significance of DET reactions of halogenated molecules for O$_3$ depletion in the polar stratosphere [5]. Oum et al. [58] had also reported that Cl$^-$ ions can be converted into Cl$_2$ molecules in atmospheric reactions of sea salts. Lu and Madey [5] therefore proposed the observed large enhancement of anions (Cl$^-$) from DET reactions of halogenated molecules adsorbed on PSC ice surfaces as an unrecognized mechanism for the formation of the O$_3$ hole. It was proposed that resultant Cl$^-$ ions can either be rapidly converted to reactive Cl atoms to destroy O$_3$ molecules, or react with other species at PSC ice surfaces to release photoactive Cl$_2$ and ClNO$_2$ in the winter (dark) polar stratosphere [5, 15]. The latter can also produce Cl atoms to destroy O$_3$, upon photolysis in the spring polar stratosphere.

Subsequently, numerous data from field measurements of total O$_3$, CFCs, CRs as well as O$_3$-loss induced stratospheric cooling over Antarctica over the past five decades were examined by Lu and Sanche [6] and Lu [14, 15]. These data have provided strong evidence of the cosmic-ray-driven electron-reaction (CRE) mechanism for the O$_3$ hole. In particular, ozone loss has shown strong spatial and time correlations with CR intensity. The electron production rate by CRs has a maximum at an altitude of around 18 km in the lower polar stratosphere, at which the O$_3$ hole is exactly observed. More remarkably, observed data have shown an *11-year cyclic* variation of polar O$_3$ loss, corresponding to the 11-year cycles of CR intensity. This is consistent with the prediction of the CRE mechanism, which is strikingly different from various photochemical model calculations predicting no 11-year cyclic variations in polar O$_3$ loss [27, 28]. It should be noted that because the oscillation amplitude of the CR intensity in 11-year CR cycles was well-known to be small, only about 10% of its mean value, the resultant oscillation amplitude of polar O$_3$ would be too small (far less than 5%) to observe if the CRE mechanism only played a minor role [14, 15].

In addition to their well-known role in O$_3$ depletion, CFCs are also long known effective greenhouse (GH) gases [59-65]. These previous studies using various climate models unfortunately concluded that halocarbons would play an important but not dominant role in past and future surface temperature changes. In current IPCC climate models [66, 67, 27, 28], it was generally thought that halocarbons would play only a minor part in global warming, whose concentrations are orders of magnitude smaller than those of non-halogenated gases (CO$_2$, CH$_4$ and N$_2$O). The 2011 WMO Report [28] has concluded that the positive radiative forcing $\Delta F$ due to the CFCs and HCFCs in 2008 was 0.34 ± 0.03 W/m², which represented only ~17% of the calculated $\Delta F$ of +1.7 W/m$^2$ by CO$_2$, together with a small $\Delta F$ of about -0.05 ± 0.1 W/m$^2$ due to stratospheric O$_3$ depletion. For a GH gas, the calculation of the radiative force change $\Delta F$ as a function of changing concentration can be simplified into an algebraic formulation specific to the gas. In climate models mentioned above, a *linear* dependence of the $\Delta F$ with concentration has been used for halocarbons, whereas a *logarithmic* relationship has been assumed for CO$_2$ so that increased concentrations have a progressively smaller warming effect. Namely, the simplified expression for the radiative force $\Delta F$ of CO$_2$ is:

$$\Delta F = 5.35 \times \ln(C/C_0) \qquad (2)$$

in Wm$^{-2}$, where $C$ is the CO$_2$ concentration in parts per million (ppm) by volume and $C_0$ is the reference concentration. With the CO$_2$ concentration rising from 285 ppm in 1850 (pre-industry era) to ~390 ppm by 2010, Eq. 2 simply gave the radiative forcing of ~1.7 W/m² for CO$_2$ alone [28]. It was thus concluded that CO$_2$ would play the dominant role in recent global warming. However, this conclusion would be valid *only if* the logarithmic relationship in Eq. 2 for the GH effect of non-halogen gases holds, i.e., there is no saturation in the GH effect.

Observed data have showed that variations of both total ozone and temperature in the lower stratosphere over Antarctica since 1950 have been predominantly controlled by CFCs and CRs and the global surface temperature change has been well correlated with atmospheric CFCs [15]. Observed data have also strongly indicated that the warming effect of CO$_2$ and other non-CFC gases has been saturated, where the saturation is defined as the phenomenon of no considerable temperature rise observed with an increase in gas concentration [68]. The actual role of halocarbons in global climate change was therefore re-evaluated. The results have shown that halocarbons alone could indeed account for the observed global surface temperature rise of ~0.6 °C from 1950 to 2002 and global temperature will reverse slowly with the projected decrease of CFCs in coming decades. Consistently, the 2011 WMO Report [28] stated that "*There have been no significant long-term trends in global-mean lower stratospheric temperatures since about 1995*".

In a recent paper, Grooß and Müller [69] stated that "Here, we scrutinise the new theories and concepts put forward by Lu, …We find that the predictions of the future development of the ozone hole based on the CRE mechanism are not reliable and that the prediction of global cooling until 2050 is incorrect". They criticized that the CRE mechanism for the $O_3$ hole and the CFC mechanism for global climate change "do not have a physical basis". To support their criticisms and to show "no evidence of the CRE mechanism", they showed correlations of CFC-12, $N_2O$ and $CH_4$ "satellite data" claimed to obtain from the Canadian ACE-FTS databases.

In contrast, Revadekar and Patil [70] have most recently used real observed data to study the effect of CFCs on surface temperature changes over the region of India during the period of 1992-2007. They applied a straightforward statistical analysis (the Pearson Correlation analysis method) to real observed data and showed that space-time distribution of correlation coefficients between CFCs and monthly temperature exhibits positive results in most of the months except the pre-monsoon (MAM) months. The observed total level of CFCs in all India increased from 1992 to 1997, and has been decreasing continuously after 1997. During increasing phase of CFCs, temperature showed increasing trends except MAM, while similar features seen but trend magnitudes decreased during decreasing phase. Based directly on real observed data, they have concluded that indeed the variation in surface air temperature has a certain link with the changes in CFCs over Indian region.

The rest of this paper is organized as follows. Section 2 will show that there are serious problems with the "ACE-FTS data" reported by Grooß and Müller [69, 71]; instead *real* data from the NASA UARS and ESA satellites will be presented. Section 3 will show that the criticisms by Grooß and Müller cannot stand scientifically. Finally, the conclusions will be given in Section 4.

## 2. On the satellite coverage and analysis of CFCs, $N_2O$, and $CH_4$ data

### 2.1 "ACE-FTS data" by Grooß and Müller

Fig. 1 of Grooß and Müller [69] showed "CFC-12/$CH_4$ correlation for different latitude regions from ACE-FTS. … panel c shows Southern polar latitudes in *February* and *March*, and panel d shows Southern polar latitudes in *October* and *November*." In the text of [69], Grooß and Müller also stated that "Fig. 1 shows the correlation of CFC-12 ($CCl_2F_2$) data with $CH_4$ … and the polar correlation for the beginning and end of the polar winter, … no significant change of the tracer correlation is observed over the Antarctic winter and spring, when PSCs are ubiquitous in the Antarctic polar vortex (Fig. 1, panels c and d). Such behaviour is in clear contrast to the predictions of DEA induced loss of CFC-12 on PSC surfaces".

Müller and Grooß then referred to their earlier paper [71] and mentioned their "demonstration" that a decomposition of CFC-12 by the DEA mechanism could not be significant by showing the year-to-year deviation from the polar tracer correlation of ACE-FTS $N_2O$ and CFC-12 data. In Fig. 2 of [71], they presented *eight time-series monthly-averaged data points per year* for each $N_2O$ and CFC-12 from the "ACE-FTS satellite data" in the polar stratosphere "poleward of 60° S". Reading the paper carefully, one can find that their use of the phase "poleward of 60° S" was actually referred to the Antarctic vortex with latitudes of 60-90° S throughout the paper (their text, Fig. 1 and references) [71] and the criticized figure of an earlier paper by Lu [14]. The debate was obviously centered on the CRE mechanism for the $O_3$ hole *in the Antarctic vortex (*60-90°S).

To reveal the serious problems with the "ACE-FTS data" reported in Fig. 1 of Grooß and Müller [69] and Fig. 2 of Müller and Grooß [71], the latitude coverage of the ACE-FTS onboard the Canadian SCISAT-1 satellite is shown in Fig. 1 here. This plot clearly shows that the SCISAT (ACE-FTS) satellite has only covered higher latitudes up to ~82° S over Antarctica in two short periods (March 20—April 10 and August 20—September 10); for other months, the satellite has essentially not covered the Antarctic vortex but the latitudes below ~63° S or the North hemisphere only. For example, obviously the satellite has not covered the Antarctic region or even the mid-latitudes of the South hemisphere in the months of *February* and *October* and has little coverage to the Antarctic vortex (60°-90°S) in *November*. Also, it is obviously impossible to obtain *eight monthly mean data* of CFC-12 and $N_2O$ in *the Antarctic vortex* from the ACE-FTS database. Note that one cannot take the data in the polar edge (60-63° S) which is not covered by the $O_3$ hole in most time to represent the average data for $O_3$-depleting reactions in the polar vortex. *Therefore, one can conclude that the "ACE-FTS data" reported in panels c and d of Fig. 1 in Grooß and Müller [69] and in Fig. 2 of Müller and Grooß [71] are obviously questionable because of the limitation in the latitude coverage of the satellite.*

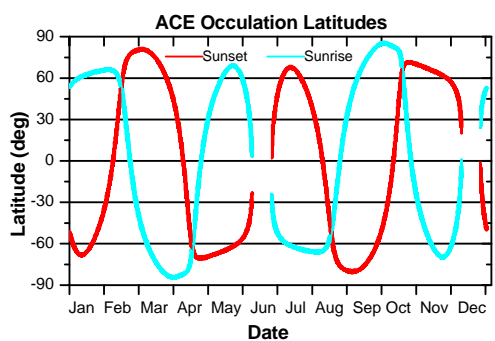

**Fig. 1.** Latitude coverage of the ACE-FTS onboard the Canadian SCISAT-1 satellite. This plot is available at the University of Waterloo's ACE website (http://www.ace.uwaterloo.ca/mission_orbit.html).

Even assuming that the September, October and November data for the Antarctic vortex were available in the ACE-FTS databases, these months are *not* in the season of *winter* but are in the spring in Antarctica. By *September* each year, the largest annual ozone $O_3$ has appeared and severe air descending has taken place and resulted in significant decreases of all CFCs and trace molecules in the polar vortex. To test the DET reaction mechanism of halogenated molecules, which is most effective in the winter polar stratosphere, one would have to look in the continuous time-series data *in the winter months (from May, June, July to August)*. This issue had indeed been pointed out previously [15]. Ignoring this key point, Grooß and Müller [69] presented not only problematic "ACE-FTS" data in their Fig. 1 but also the two-month data of $CH_4$, $N_2O$ and CFC-12 in *March* and *September* in their Fig. 2 to show "no obvious evidence" of the DET reactions of CFC or $N_2O$ over Antarctica during *winter*. Even the fits to the data in their Fig. 1c and d resulted in no *linear* correlations, which should exclude the air mixing process as the cause of concentration changes, according to their earlier argument [72].

**2.2 Data in the references by Grooß and Müller**

Grooß and Müller [69] also cited references to argue that "Very compact correlations are reported in the literature between observations of long-lived tracers like $CH_4$ and $N_2O$ and CFCs (e.g., Michelsen et al., 1998; Müller et al., 2001)." Looking in the cited references, however, one can find that Michelsen et al. [73] actually reported data obtained from the ATMOS instrument for the late springtime (namely in the period of *November 4-12, 1994*) but not winter polar stratosphere. Müller et al. [72] did report the balloon-borne measured data of $O_3$, $N_2O$ and CFCs over the 1991/92 Arctic winter (December 1991, January, February and March 1992). In contrast to the later claim and argument by Grooß and Müller [69], however, Müller et al. [72] did show an excellent linear correlation for their $N_2O$-CFC12 data (plotted as Fig. 4 in [72]) but not for their $N_2O$-CFC11 and $N_2O$-CFC113 data. Interestingly, Müller et al. [72] even concluded that "Analysis of the concurrently measured CFC11-$N_2O$ relation rules out mixing processes as the major cause" and "chemical loss is the reason" for the observed $O_3$ deficit of about ~25% in the lower Arctic stratosphere in the dark middle winter of 1992. They also stated that "No further chemical ozone loss is observed between early February and late March 1992" in the Arctic vortex. These observations cannot be explained by the photochemical model for polar $O_3$ loss, as Müller and co-workers [74] concluded "The model significantly underestimates the loss rates inferred for January to mid-February. Extensive sensitivity studies show that the discrepancy between model and Match results cannot be explained by the known uncertainties in the model parameters."

**2.3 Arguments by Grooß and Müller**
Besides the problems with their "ACE-FTS data", Müller and Grooß [69, 71] also presumed that no physical or chemical processes could occur for $N_2O$ in the lower polar stratosphere because no photo-dissociation of $N_2O$ (the so-called 'trace gas') occurs there. They argued that "a DEA-induced decomposition of $N_2O$ or $ClONO_2$ is speculative, since the DEA mechanism for stratospherically relevant species has only been demonstrated in the laboratory for CFC-11, CFC-12, and HCl". This argument is strange. First, $N_2O$ is a very well-known and widely-used scavenger for electrons produced in radiolysis of water in radiation biology [75] and its DET reactions with weakly-bound electrons at various surfaces have been well documented (see references in [15]). Second, as reviewed recently [15], it has been well established both experimentally and theoretically that for those molecules such as Cl-, Br-, and I-containing molecules that are effective for dissociative attachment to free electrons at energies near zero eV in the gas phase, their DET reactions with weakly-bound electrons are highly effective in polar media such as $H_2O$ and $NH_3$ in various phases [5, 6, 36-57]. Particularly, gaseous $ClONO_2$ has been reported to have an extremely large DEA cross section at an electron energy of zero eV. Therefore, it is certainly reasonable to conclude that similar to CFCs, $ClONO_2$ and $N_2O$ will have effective DET reactions with weakly-bound electrons trapped in PSC ice under the strongest CR radiation in the winter polar stratosphere at altitudes of 15-20 km, though their DET reaction efficiencies (absolute cross sections) are expected not to be identical [15]. $CH_4$ should in principle be an inerter "trace" gas than $N_2O$/CFCs, because $CH_4$ does not have a dissociative attachment resonance with nearly 0 eV electrons in the gas phase and therefore its direct DET reaction with weakly-bound electrons trapped in PSC ice cannot occur. Nevertheless, it should be cautious to assume an *absolute* inertness of $CH_4$ in the stratosphere as there is also evidence of the potential reaction of $CH_4$ with reactive radicals produced from the reactions of other molecules such as $N_2O$ and CFCs [15]. In fact, the reaction $CH_4$ with a Cl atom arising from photolysis of CFCs had been included in photochemical models to form HCl [22-25].

**2.4 Data from NASA UARS and ESA satellites**

In fact, continuous, time-series data of CFC-12 and $CH_4$ over the Antarctic vortex during winter months (June, July and August) are available in NASA and ESA satellite databases. In Fig. 19 of Lu [15], one set of such time-series data was presented: the observed monthly average $CH_4$ data combining the data from both the Cryogenic Limb Array Etalon Spectrometer (CLAES, running over the complete year 1992 only) and Halogen Occultation Experiment (HALOE, running over six years in 1992–1998) aboard the NASA Upper Atmosphere Research Satellite (UARS), and the $CF_2Cl_2$ data over the winter of 1992 from the CLAES were shown. The data clearly indicate that DET reactions of CFCs but not $CH_4$ occurred in the lower polar stratosphere during the whole period of winter [15]. Grooß and Müller [69] criticized that the data for $CH_4$ from the two instruments (HALOE and CLAES) aboard the UARS would not be comparable with the data of CFC-12 from the CLAES only. There are several problems with this criticism. First, the combined $CH_4$ data from the instruments of CLEAS and HALOE were processed and made available to researchers directly by the HALOE team rather than by Lu [15]. Second, the combined HALOE-CLAES $CH_4$ data were used in Lu [15] just because the multiple-year data are more reliable than the CLEAS $CH_4$ data that covered the single year 1992 only. Note that there are no combined HALOE and CLAES CFC-12 data available. Third, time-series $CH_4$ and CFC-12 data from the same instrument CLEAS are now plotted in Fig. 2 here, which shows a consistent result with Fig. 19 of [15]. These data clearly show that for $CH_4$, no decrease in the *lower* Antarctic stratosphere *below 20 km* from March to August (even an increase in July) was observed, whereas the $CF_2Cl_2$ level significantly and continuously decreased from ~320 in March to 200 pptv in August. These data provide strong evidence of DET reactions of CFC-12 but not $CH_4$ in the *winter* lower polar stratosphere.

Although the CLEAS data covered one full year (1992) only and might have large uncertainties, new and more reliable satellite data over the past decade are now available. Here, original ESA's Oxford MIPAS Near Real Time satellite data of CFC-12, $N_2O$ and $CH_4$ in the lower Antarctic stratosphere (65-90° S) during the winter season (June 23-September 30) *over the past ten years* (2002-2011) are plotted in Fig. 3A-C. The 9-year (2003-2011) mean time-series data are obtained by averaging these satellite data and shown in Fig. 3E, where the CFC-12/$N_2O$/$CH_4$ data were normalized to its value in the beginning of winter. In the averages, the 2002 data were not included due to the unusual split of the polar vortex in 2002 [27]. Interestingly, Fig. 3E shows that the CFC-12 data exhibit a variation curve that overlaps well with that of the $N_2O$ data. In contrast, the $CH_4$ data show a very different curve. It is clearly confirmed that the CFC-12 and $N_2O$ levels exhibit a similar continuous decrease since the beginning of winter, while the $CH_4$ level does not decrease until the end of August. After that, all gases show decreasing trends in September-October and then rising trends in November. Note that in September and October, the levels of all gases ($CH_4$, $N_2O$ and CFC-12) drop in the polar lower stratosphere. This can be well explained by significant stratospheric cooling and air descending as a result of severe ozone loss in the spring lower polar stratosphere. These data provide strong evidence of the CRE reactions of CFCs and $N_2O$ in the *winter* polar stratosphere.

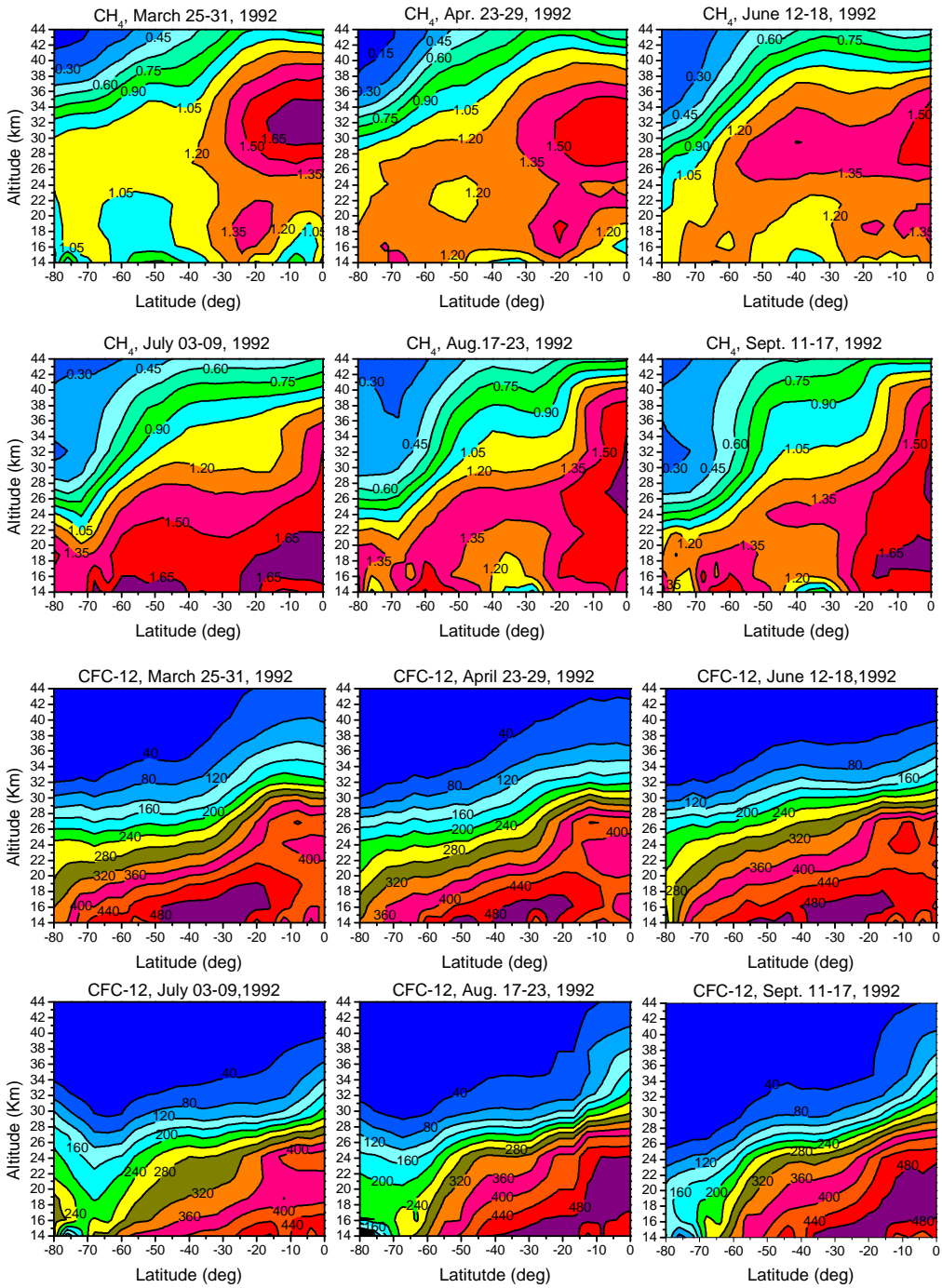

**Fig. 2.** Weekly averaged zonal-mean altitude-latitude maps of $CH_4$ and CFC-12 mixing ratio in the southern hemisphere from March to September 1992. Data (V9) were obtained from the CLAES aboard the NASA Upper Atmosphere Research Satellite (UARS).

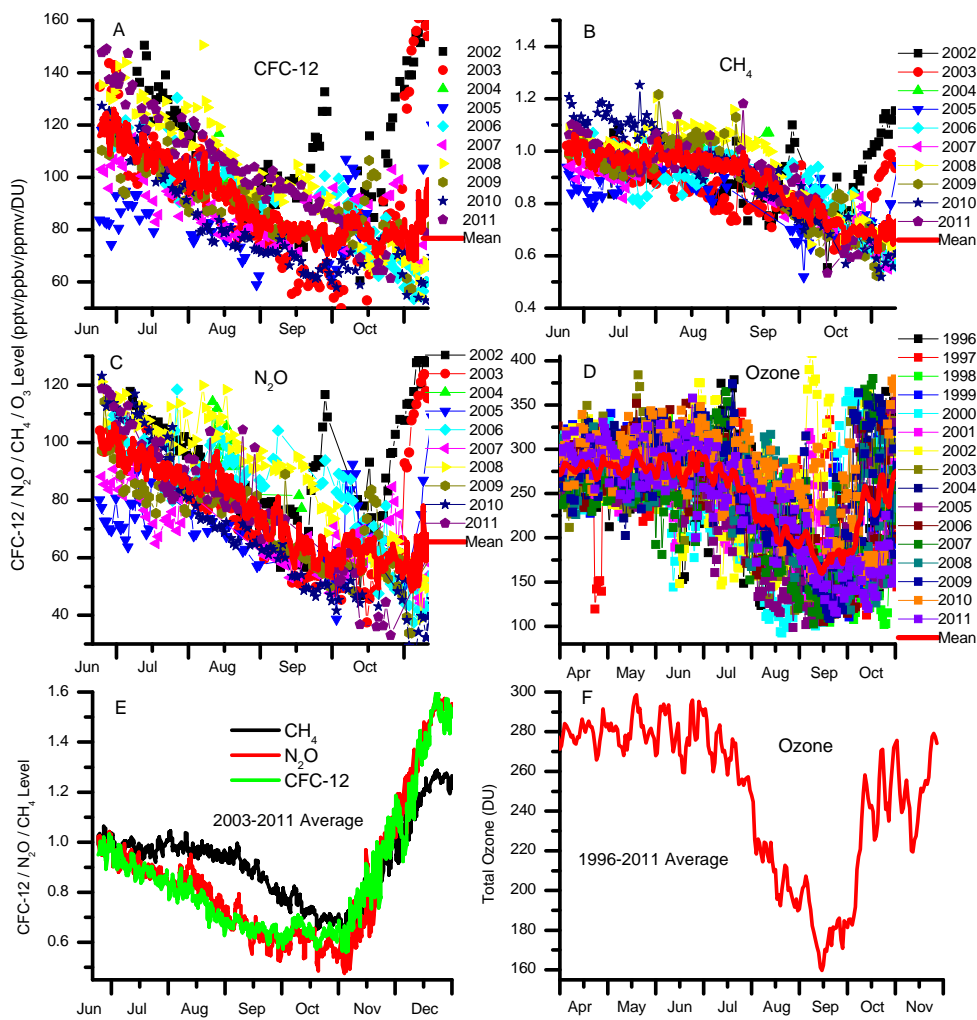

**Fig. 3. A-C:** ESA's Oxford MIPAS Near Real Time satellite data (symbols) of CFC-12, $N_2O$ and $CH_4$ in the lower (30 mb) Antarctic stratosphere (65-90° S) during the winter season (June 23-September 30) over the past ten years (2002-2011). The solid line in red in each plot is the 9-year (2003-2011) mean time-series data obtained by averaging original data. **D.** British Antarctic Survey (BAS)'s real-time daily total ozone data measured at Rothera, Antarctica over the past 16 years (1996-2011). The solid line in red is the 15-year mean time-series data, obtained by averaging original data with the 2002 data excluded. E: The 9-year mean ESA's Oxford MIPAS time-series data of CFC-12, $N_2O$, $CH_4$, where the data for each gas were normalized to its initial value in the beginning of winter. F: The 15-year mean BAS time series data of total $O_3$ at Rothera. In E and F, the 2002 data were not included due to the unusual split of the polar vortex [27].

As mentioned in the above, the CRE mechanism can form reactive species to destroy ozone in both the *winter* polar stratosphere in the dark and the *springtime* polar stratosphere with sunlight. Fortunately, real-time daily total ozone variations at Rothera, Antarctica over the whole years have been well recorded by the British Antarctic Survey (BAS) since 1996. The original total ozone data measured at Rothera over the past 16 years (1996-2011) are plotted in Fig. 3D. Fig. 3F shows the 15-year mean time series total ozone data averaged from these data, excluding the 2002 data due to the unusual split of the polar vortex in 2002. It is clearly seen that total $O_3$ starts to drop from a high value of about 300 DU at the beginning of July to about 220 DU at the middle of August and to a minimum value as low as 150 DU in September. Although the ozone hole, which is currently defined in ozone research community as the area with total $O_3$ below 220 DU, would only appear after the middle of August, significant $O_3$ loss has occurred far before the mid-August in the winter polar stratosphere. Note that there is lack of sunlight in the lower polar stratosphere during winter. Thus, the significant polar $O_3$ loss in July and early August cannot be explained by the photochemical models. The real-time variation of total $O_3$ over Rothera (Fig. 3F) generally follows well that of CFCs or $N_2O$ (Fig. 3E) during winter, indicating that the CRE mechanism plays an important role in causing severe $O_3$ loss over Antarctica. Note that if N$_2$O is used as a 'trace' gas and its concentration drop in early and middle winter is used to 'correct' total O$_3$ data measured by satellites, then 'no chemical ozone loss' prior to the mid-August would result in. This 'calibration', however, would inevitably give rise to artificial results.

So far, no real evidence for the absence of DET reactions of halogenated molecules in the polar stratosphere *during the winter months* has been found. *Instead, real, observed time-series data from multiple satellites have provided strong evidence of DET reactions in the winter polar stratosphere, as shown in Fig. 19 of [15] and Figs. 2 and 3 in the present paper.*

## 3. On the conclusions for stratospheric ozone and global climate

Grooß and Müller [69] made further untrue criticisms. They criticized that the CRE mechanism [15] and the CFC mechanism [68] are "based only on correlations of two parameters" and "do not have a physical basis", "which can only give an indication of a cause-and-effect chain, and therefore there are no proofs of the theories". These criticisms do not agree with the facts mentioned in the Introduction. In fact, the CRE mechanism was developed from solid observations of DET reactions of CFCs on ice surfaces from laboratory measurements [5, 36-52]. Obviously, some of these observations had been made far before strong correlations between O$_3$ loss and CR intensity and between temperature and CFCs were discovered [6, 14, 15]. Obviously, the criticisms of Grooß and Müller [69] ignored the basic fact: it is the substantial data from laboratory measurements, which were actually reviewed in the larger first half of the whole paper [15], that led to the findings of the CRE mechanism and the correlations in data from field observations. The data from both laboratory and field measurements have provided strong evidence of the CRE mechanism for the ozone hole. And there exists a solid physical basis for the CFC mechanism of global climate change [59-65, 68, 70]. There are also many invalid criticisms raised by Grooß and Müller [69]. These will be revealed point-by-point below.

### 3.1 Relationship between O$_3$ depletion and PSCs

Grooß and Müller [69] argued against a statement that "the ozone hole is exactly located in the polar stratosphere at ~18 km, where the rate of ionization for cosmic rays producing electrons shows a maximum". First, it should be noted that this is an observational description [6, 14, 15], rather than an argument under debate, though one might have a different explanation about the observed fact. Grooß and Müller also made an obviously contradictory argument that "the current understanding is that the vertical extent of the ozone loss in the ozone hole is determined by the temperature structure of the polar stratosphere. Heterogeneous chlorine activation as a prerequisite for ozone depletion occurs at low temperatures on different stratospheric particles, PSCs." On the other hand, they argued that "In Arctic winter, when the occurrence of ice PSCs is rare, ozone depletion has also been reported, which cannot be explained by the CRE mechanism". In fact, the formation of PSCs is important for both the CRE mechanism and the heterogeneous reaction mechanism, and it has been well observed in recent years that O$_3$ loss over Antarctica and Arctic is very sensitive to the formation and density of PSCs. This observation, as a matter of fact, agrees with the CRE mechanism better than the photochemical model for the following reason. In the latter model, heterogeneous surface reactions of ClONO$_2$ and HCl on PSC ice lead to the formation of photoactive Cl$_2$ and HOCl in the *winter* polar stratosphere, followed by gas-phase photodissociaton of these species into Cl atoms to destroy O$_3$ in the *springtime* polar stratosphere. However, the observed fact is that the concentrations of ClONO$_2$ and HCl *rapidly* drop to very low, nearly zero levels even in the very early period (June-July) of the Antarctic winter [76]. This has been constantly observed, essentially independent of warmer or colder winters and larger or smaller Antarctic O$_3$ holes over the past two decades. The photodissociation of Cl$_2$ and HOCl in *the gas phase* would not be sensitive to the polar stratospheric temperature, definitely not be more effective in a colder stratosphere. In contrast, the observed CFC-12 shows a continuous decrease from the beginning to the end of the winter season, which is reduced to a nonzero level even at the end of winter, and the amount of its reaction through the CRE mechanism is certainly sensitive to the formation and density of PSCs during the whole winter. Note again that DET reactions of halogenated molecules have been observed in gas-, liquid- and solid-phase polar media such as H$_2$O and NH$_3$, but their reaction efficiencies are different in various phases and therefore sensitive to temperatures [15].

### 3.2 Quantitative understanding of polar O$_3$ loss and predictions of the future O$_3$ hole

The proposed CRE mechanism was used to deduce an analytic CRE equation that reproduced total $O_3$ data over Antarctica since the 1950s and made quantitative predictions of the Antarctic $O_3$ hole in the future [15]. The total ozone ($[O_3]_i$) in the polar stratosphere is given by:

$$[O_3]_i = [O_3]_0 \times [1 - k \times EESC_i \times I_i \times I_{i-1}], \qquad (3)$$

where $I_i$ is the CR intensity in the $i$th year, $EESC_i$ is the equivalent effective stratospheric chlorine, $[O_3]_0$ the total $O_3$ in the polar stratosphere when there were no CFCs in the atmosphere and $k$ a constant [15]. It has been shown that Eq. 3 with CR intensity and EEFC as sole variables reproduces well observed 11-year cyclic variations of both total $O_3$ and stratospheric temperature in the Antarctica $O_3$ hole over the past five decades (see Figs. 16a, 17b and 20c in [15]). Facing these convincing results, Grooß and Müller [69] criticized that "This fit to EESC and the 11-year solar cycle-based CR modulation (Fig. 20c of Lu, 2010a) is not able to reproduce the ozone data, e.g. around 1990. This is expected since the fit is not based on the mechanisms responsible for ozone depletion and therefore is not suited for making predictions". First, this criticism obviously contradicts with their earlier argument "For recent decades, the detection of the solar-cycle signal has been complicated by the fact that the eruptions of El Chichón (1982) and Mt. Pinatubo (1991) occurred near the declining phases of the solar cycle in 1982–1984 and 1992–1994, so that there is the possibility of aliasing volcanic and solar-cycle related effects on column $O_3$" [71]. Climate models and observations did show that a major volcanic eruption such as Mt. Pinatubo in 1991 causing a large negative radiative forcing of roughly $-3$ W/m$^2$ can lead to a large impact on stratospheric $O_3$ depletion and climate that persists for a few years [28]. Second, the results reported in Figs. 20c and 17b of Lu [15] are now updated as Fig. 4A and 4B here. It is clearly shown that in striking contrast to Grooß and Müller's criticism, the CRE model well reproduces not only observed total $O_3$ data but also ozone-loss-induced stratospheric cooling data, albeit large fluctuations of observed $O_3$ / temperature data in some years (e.g., around 1990 and 2002). It was indeed reasonable to conclude that "Although atmospheric dynamics and meteorological conditions could cause large total $O_3$ fluctuations from year to year, a long-term trend of the polar $O_3$ loss is predictable" [15]. These results have shown strong evidence and the dominance of the CRE mechanism. In contrast, the calculations using various photochemical models have been unable to either reproduce or predict 11-year cyclic total $O_3$ variations in the ozone hole [27, 28]. According to Grooß and Müller [69], the failing is due to "inter-annual variability of the atmosphere prevents accurate predictions for a particular year from simulations. Furthermore, climate-change-driven changes in the stratosphere, e.g., in stratospheric temperatures and in stratospheric circulation, will have an important impact on future stratospheric ozone levels."

Besides, the October average zonal-mean total $O_3$ data over Antarctica (60-90°S) were directly calculated by averaging the original NASA data over 5° latitude bands without area weighted in Lu [14, 15]. Those total $O_3$ absolute values are ~30 DU lower than the values calculated with area-weighted. However, the observed total $O_3$ data were presented as relative values (percents) in Lu [14, 15], and therefore the difference had almost no influence on the conclusion. As shown in Fig. 4A here, re-calculations using area-weighted monthly average zonal-mean total $O_3$ data give a good fit to Eq. 3, similar to Fig. 20c of [15]. Therefore, the criticism by Grooß and Müller [69] that predictions of future polar total $O_3$ values based on Eq. 3 "cannot be considered meaningful" is clearly invalid.

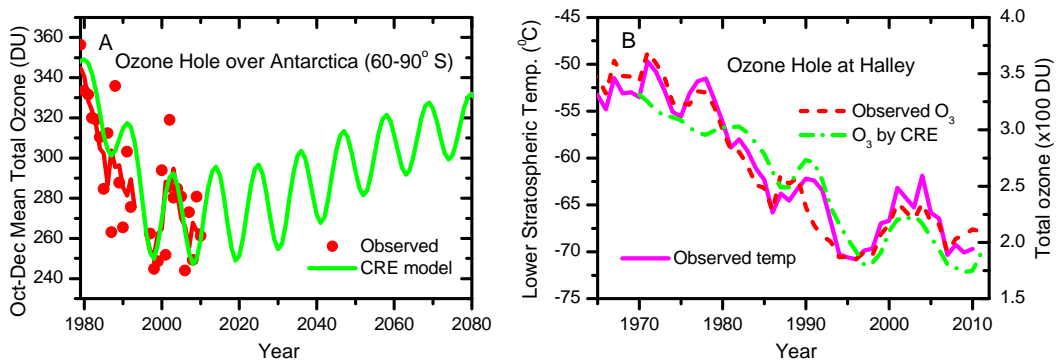

**Fig. 4.** A: Observed and calculated 3-month (October-December) average total O3 over Antarctica with latitudes (60-90o S) in 1979-2010. Calculated data (solid line in green) are obtained by the CRE equation [15]. Observed data (solid circles in red) are obtained from zonal-mean total O3 data in NASA N7/M3/EP/OMI satellite datasets with *area weighted* (modified and updated from Fig. 20c of [15]). B. Observed temperatures at the lower stratosphere (100 hPa) and total O3 over Halley (75°35' S, 26°36' W) in November (after the O3 hole peak) as well as calculated total O3 data by the CRE equation from 1965 to 2010, where a 3-point average smoothing was applied to observed data (updated from Fig. 17b of [15] with the 2002 data excluded).

**3.3 Photochemical model simulations**
Grooß and Müller [69] argued that (photo)chemistry-climate models (CCMs) for polar ozone loss do not use *observed* temperatures and winds and can therefore only calculate temperatures and ozone in a climatological sense. Therefore, "statements about particular years of CCM results are not meaningful". Chemistry transport models (CTMs), in which simulations are based on *observed* temperatures and winds, allow the observed ozone to be reproduced much better. One can find that this argument, in fact, acknowledges that current photochemical models including CTMs do not have the capability to *predict* future changes of polar ozone because they require using *observed* temperatures and winds. Indeed, researchers [29] have recently concluded in studying the Arctic ozone hole that "current climate models do not fully capture either the observed short-timescale patterns of Arctic variability or the full extent of the observed longer-term cooling trend in cold stratospheric winters; nor do they agree on future circulation changes that affect trends in transport. Our ability to predict when conditions similar to, or more extreme than, those in 2011 may be realized is thus very limited. Improving our predictive capabilities for Arctic ozone loss, especially while anthropogenic halogen levels remain high, is one of the greatest challenges in polar ozone research". This is drastically different from the capability of the CRE mechanism, which not only reproduces observed total ozone data and ozone-loss-induced stratospheric cooling data but also gives a direct prediction of the future change of the polar ozone hole, as simply expressed in Eq. 3.

**3.4 Correlation of stratospheric temperature with total $O_3$ over Antarctica**
The statement of Grooß and Müller [69] that "Thus, the correlations between temperature and ozone column shown in Fig. 17 and 18 of Lu (2010a) are caused by the radiative effect of the reduced ozone concentrations in the ozone hole" is essentially the same as the statement in Lu [15]: "Less ozone in the stratosphere implies less absorption of solar and infra-red radiation there and hence a cooler stratosphere." Thus, Grooß and Müller didn't provide any new insight into this point but just repeated the discussion that had been made by Lu [15]. Moreover, Grooß and Müller's final mentioning that "The impact of the solar cycle on stratospheric temperatures has also been investigated, the maximum effect is in the order of a 1K response amplitude located in the *tropical upper* stratosphere (…)." is misleading or irrelevant. First, the direct solar effect is not expected to affect either total $O_3$ or temperature in the *lower polar* stratosphere [27, 28]. This is indeed proven by the observed data of summer total $O_3$ in Fig. 16c of Lu [15]. Second, although the stratospheric cooling effect of $O_3$ depletion had been well observed [77, 78, 27], Lu [15] reported *the first* observation of *11-year cyclic* stratospheric cooling caused by polar $O_3$ loss (see Fig. 4B) and concluded that "These data (Figs. 17 and 18) clearly demonstrate that the long-term temperature change in the lower polar stratosphere is solely dependent on the variation of total ozone; both exhibit pronounced 11-year cyclic variations controlled by the levels of CFCs and cosmic rays. This is clearly in contrast to the previous attribution of the non-monotonic decrease in the stratospheric temperature to the effect of volcanic eruptions […] or to the effects of non-CFC greenhouse gases […]. The present observations provide strong evidence that the CRE-driven polar ozone loss results in a subsequent 11-year cyclic stratospheric cooling."

**3.5 CFCs and climate change**
Ignoring nearly all of the observed and theoretical results presented in Figs. 17, 18, 21 and 22 of Lu [15] and Figs. 1-6 of Lu [68], Grooß and Müller [69] re-plotted *only part of* the data presented in the *single* Fig. 21 of Lu [15] and criticized Lu's conclusion that CFCs, rather than $CO_2$, were the major culprit for global warming in late 20th century. They argued that "The inspection of this figure (in the same way as Fig. 21 of [15] is interpreted) could suggest that surface temperatures are equally well (or perhaps better) correlated with $CO_2$ concentrations than with EESC. ….". It must be pointed out that Grooß and Müller [69] quote out of context on Lu's results for the relationship between CFCs and global climate change, which was based on a large amount of data and discussion presented in Lu [15, 68]. For example, Fig. 2 of Lu [68] showed measured data of global surface temperature versus $CO_2$ concentration over the eight decades from 1850 to 1930 (when there were no CFCs emitted into the atmosphere), which results in a nearly zero correlation coefficient

R (=0.02). If the method of Müller and Grooß [71] using a low linear correlation coefficient (≤0.5) fitted from their $O_3$ data with only 16 data points and large fluctuations to exclude the CRE mechanism is followed, then the $CO_2$ global warming model would have been ruled out much earlier with no need of any other evidence. Furthermore, Fig. 3 of Lu [68] showed time-series data of atmospheric $CO_2$ concentration, the total concentration of major atmospheric halocarbons (CFCs, $CCl_4$ and HCFCs) and global surface temperature over nearly six decades from 1950 to 2009. It is seen that global surface temperature was nearly constant from 1950 to ~1975, then had a linear rise from 1975 to ~2002 and finally turned to a slow decrease after 2002. Most strikingly, it is shown that global surface temperature has had a nearly perfect linear correlation with the total loading of halocarbons in the atmosphere from the 1950 to the present; *the linear fit gives a statistical correlation coefficient R as high as 0.96*. In fact, consistent with Lu's observations, the 2011 WMO Report [28] states that "New analyses of both satellite and radiosonde data give increased confidence in changes in stratospheric temperatures between 1980 and 2009. The global-mean lower stratosphere cooled by 1–2 K and the upper stratosphere cooled by 4–6 K between 1980 and 1995. *There have been no significant long-term trends in global-mean lower stratospheric temperatures since about 1995*."

The observed data mentioned above strongly indicate that global temperature is highly sensitive to the variation of atmospheric halocarbons rather than increasing $CO_2$ and other non-halogen GH gases. In current climate models [66, 67, 27, 28], the radiative force of $CO_2$ can approximately be calculated by Eq. 2, implying that the global temperature sensitivity would *decrease* with increased $CO_2$ concentrations. This actually contradicts the observed temperature variations of about 0.2 °C during 1850-1970 and of ~0.6 °C during 1970-2002, not to mention the observed *negative* correlation with $CO_2$ at higher concentrations (≥373 ppm) since 2002. In contrast, global temperature has exhibited a nearly perfect linear positive correlation (R=0.96) with the total concentration of halocarbons since their considerable emission into the atmosphere in the 1970s. All these data strongly indicate that the greenhouse effect of increasing non-halogen gases has been saturated and that halocarbons (mainly CFCs) have played the dominant role in global climate change since 1970.

Grooß and Müller [69] also made an incorrect argument against the work of Lu [68] by stating "He argues that CFCs show a larger greenhouse effect than $CO_2$ on the basis of a simplified radiation model calculation, looking at transmitted radiation *at the surface* and claims that the $CO_2$ lines would be saturated." This statement severely distorts the observations reported in Lu [68]. For example, Fig. 5 of [68] showed observed radiance spectra of the Earth's outgoing longwave radiation (OLR) measured by the NASA Infrared Interferometric Spectrometer (IRIS) onboard the Nimbus 4 spacecraft in 1970 and the IMG instrument onboard the Japanese ADEOS satellite in 1997 and the theoretical OLR difference spectra. "These data strikingly show that the expected strong $CO_2$ absorption band in the 700 to 800 $cm^{-1}$ region does not appear in the observations of the difference radiance spectrum between 1970 and 1997 (…). This radiative forcing arising from the $CO_2$ absorption has lied at the heart of the debate on the anthropogenic global warming observed in the late half of the 20th century. … These observations and calculations of radiance spectra therefore also provide strong evidence of the saturation in the warming effect of $CO_2$." In fact, the strong evidence for the saturation in GH effect of non-CFC gases was drawn first from observed data of polar stratospheric temperatures over Antarctica since the 1950s (Figs. 17 and 18 of [15]), then from global surface temperatures versus $CO_2$ since 1850 and CFCs since 1970 (Figs. 21 and 22 of [15]; Figs. 2 and 3 of [68]) and theoretical and observed transmitted radiation spectra measured both at the surface and at the top of the atmosphere (Fig. 4 of [68]), as well as theoretical and measured OLR spectra by satellites at the top of the atmosphere at altitudes of about 800 km (Fig. 5 of [68]). Moreover, the elucidation of the Earth blackbody radiation feature, as presented in Fig. 1 of [68], has also given a sound physical basis for the saturated GH effect of non-halogenated gases. It was clearly elucidated that CFCs, rather than $CO_2$ or $CH_4$ / $N_2O$, have IR absorption bands centering at the blackbody radiation spectrum of the Earth or the atmospheric window at 8-12 μm [68]. Moreover, non-halocarbon gases have so high concentrations, ~390 ppm for $CO_2$, 1.5-1.8 ppm for $CH_4$ and 300-325 ppb for $N_2O$, which are respectively $10^6$, $10^4$ and $10^3$ times those of CFCs and HCFCs in 100-500 ppt, that their GH effects are saturated.

The magnitude of the GH effect depends on the sensitivity of the climate system. There exists a large uncertainty in *climate sensitivity factor α* [66, 67], which can be defined as the amount of temperature change per change in direct radiative forcing driven by GH gases, i.e. $\alpha=dT/dF$, with no climate feedback included. In current climate models, equilibrium climate sensitivity was often determined by projecting $\Delta T_{2x}$ to be a particular value corresponding to a *calculated* radiative force of $\Delta F \approx 3.7$ W/m$^2$ *calculated* by Eq. 2 arising from a doubling of atmospheric CO$_2$ concentration [66, 67]. The $\Delta T_{2x}$ value was therefore termed *climate sensitivity*. It is also known that atmospheric feedbacks largely control climate sensitivity. As global average temperature increases, tropospheric water vapor increases and this represents a key positive feedback of climate change. It has been estimated that water vapor feedback acting alone approximately doubles the warming from what it would be for fixed water vapour [67]. Moreover, water vapour feedback can also amplify other feedbacks such as cloud feedback and ice albedo feedback in models. The change in equilibrium surface temperature ($\Delta T_s$) arising from the radiative forcing $\Delta F$ of the surface-atmosphere system due solely to a GH gas increase can be calculated by: $\Delta T_s = \alpha\beta\Delta F$. Given no saturation in CO$_2$ GH effect, the current large atmospheric CO$_2$ concentration (~390 ppm) would result in a very large positive radiative force (~1.7 W/m$^2$) given by Eq. 2 [28]. Consequently, observed temperature changes have forced climate models to use a very low climate sensitivity [$\alpha\beta \leq 0.5$ K/(W/m$^2$)]. Unfortunately, this has inevitably led to the general conclusion that halocarbons are important, but by no means dominant for surface temperature changes [60-67, 27, 28].

The observed evidence of the saturated GH effect of non-halocarbon gases led to a new evaluation of the effect of halocarbons on global climate [68]. Taking zero GH effects of non-halocarbon gases, Lu [68] re-calculated the global surface temperature change $\Delta T_s$ due solely to halocarbons. As updated in Fig. 5, the calculated results with a climate sensitivity factor $\alpha=0.9$ K W$^{-1}$m$^2$ and a climate feedback amplification factor $\beta=2$ reproduce the observed temperature data excellently. It should be noted that the results of Lu's model calculations are consistent with those of previous calculations on the GH effect of halocarbons using various climate models by other researchers [59-67]. The only distinction is that unlike previous climate models [59-67], Lu's climate model calculations are based on the observed saturation in GH effect of non-halogenated gases (CO$_2$, CH$_4$ and N$_2$O) and only the GH effect of halocarbons is taken into account.

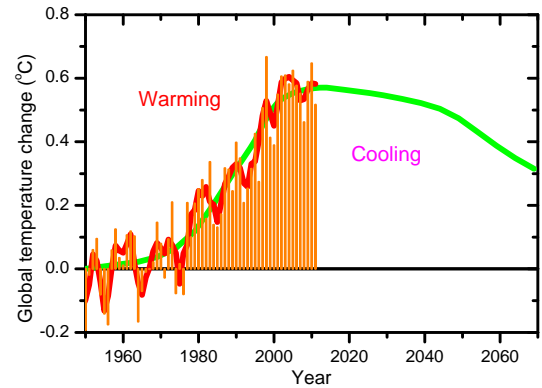

**Fig. 5.** Calculated change in equilibrium surface temperature ($\Delta T_s$) (solid curve in green) due solely to CFCs, HCFCs and CCl$_4$, with a climate sensitivity factor $\alpha=0.9$ K W$^{-1}$m$^2$ and a climate feedback amplification factor $\beta=2$. Observed global surface temperature data (bars) were from the UK Met Office Hadley Centre; the red curve is a 3-point average smoothing of observed data (updated from Fig. 6 in Lu [68]).

In contrast to the above observed facts, Grooß and Müller [69] made very confusing and contradictory statements. For example, their statement "all known projections show that the trend of global warming will continue as long as CO$_2$ increases in the atmosphere (Solomon et al., 2007b)" contradicts with "Indeed, because the recovery of the atmosphere from an anthropogenic CO$_2$ perturbation will take place on time scales of centuries or longer, the climate change that takes place due to anthropogenic CO$_2$ increases will be largely irreversible for 1000 years after emissions have stopped (…; Solomon et al., 2009)". They also stated "the globally averaged surface temperatures are controlled by a combination of multiple effects, most prominently the increase in emissions of CO$_2$ and other greenhouse gases, … (see e.g., Solomon et al., 2007b)" and "A plausible explanation for the recent decrease of surface temperatures is a combination of low El Niño index, low solar activity, and the Kasatochi volcano eruption in 2008". Reading these confusing statements, one would be very difficult to find out that in Grooß and Müller's arguments, what would the major culprit for the observed global warming in the late half of the 20$^{th}$ century be: was it due to the anthropogenic CO$_2$ or other GH gases or natural effects? One point for sure is that if there were a delay of 1000 years in the GH effect of CO$_2$, then the global warming observed for the period of 1950 to 2000 must not be due to anthropogenic CO$_2$ since the industrial revolution started at 1850 only.

## 4. Conclusions

Based on the latitude coverage of the ACE-FTS satellite, one can conclude that there are serious problems with the "ACE-FTS data" reported by Grooß and Müller [69, 71]. Therefore, their statement "By analysing the ACE-FTS data, we demonstrated that there cannot be significant CFC decomposition besides photolytic decomposition in the stratosphere by the proposed CRE effect (Lu, 2010a)" cannot stand scientifically. And a large number of data and scientific facts mentioned in this paper have disproven Grooß and Müller's fact-lacking criticisms that "the methods of analysing ozone and global temperature data used by Lu (2010a) which are based solely on correlations of parameters, are not conclusive to explain the complex processes both of ozone depletion and surface temperature development. Thus, meaningful predictions based on the correlation of EESC and temperature anomalies cannot be drawn. The strong conclusions for climate models put forward by Lu (2010a) do not have a physical basis".

On the contrary, one can find that it is numerous laboratory observations of DET reactions of halogenated molecules that led to the findings of strong spatial and time correlations between polar $O_3$ loss or stratospheric cooling and CRs and the birth of the CRE mechanism of the $O_3$ hole. Substantial observed data from both laboratory and field measurements have provided strong evidence of the CRE mechanism. The GH effect of CFCs has been known for a long time. However, the new conclusion that CFCs, rather than $CO_2$, were the major culprit for global warming in late 20th century and the prediction that a long-term global cooling starting around 2002 will continue for coming five to seven decades were made with a sound physical basis and numerous observed data. These included polar stratospheric temperatures and global surface temperatures, atmospheric transmittance spectra of the infrared radiation and outgoing longwave radiation spectra of the Earth, as well as the elucidation of the essential feature of the Earth blackbody radiation and new calculations of the GH effect of halocarbons [15, 68].

Finally, their statement in the Conclusion of [69] "The finding of the IPCC '…' (Solomon et al., 2007b) remains unchallenged by the analysis of Lu (2010a)" obviously contradicts their own statements in their Abstract and Introduction that Lu's conclusions "challenge the fundamental understanding of polar ozone loss and global climate change".

**Acknowledgements**

The author is greatly indebted to the following Science Teams for making the data available: NASA TOMS and OMI Teams for ozone data; NASA UARS's CLAES and HALOE Teams for CFCs and $CH_4$ data; The British Antarctic Survey's Ozone Team for ozone at Rothera and Halley, Antarctica and lower stratospheric temperature data at Halley (credited to Dr. J. D. Shanklin); The University of Oxford's MIPAS team (Dr. Anu Dudhia) for Near Real Time satellite data of CFC-12, $N_2O$ and $CH_4$; The Bartol Research Institute's Neutron Detection Team for cosmic ray data at McMurdo; the UK Met Office Hadley Centre for global surface temperature data; and the Canadian ACE team for providing the latitude coverage plot of the ACE-FTS satellite and for pointing out the limitations of the ACE-FTS datasets. This work is supported by the Canadian Institutes of Health Research (CIHR) and Natural Science and Engineering Research Council of Canada (NSERC).